\documentclass{appolb}
\usepackage{graphicx}

\begin{document}
\title{Photon-photon scattering in the resonance region at
  midrapidity at the LHC.%
  \thanks{Presented at Diffraction and Low-x Conference 2018,
    Reggio Calabria, \mbox{Aug $\!$26 - Sept $\!$1}}%
}
\author{Rainer Schicker
\address{Phys. Inst., Heidelberg}
\\
  {Mariola K\l{}usek-Gawenda, Antoni Szczurek}
\address{Inst. Nucl. Phys., PAN Krak$\acute{o}$w, Univ. of Rzesz$\acute{o}$w}
}

\maketitle
\begin{abstract}
A study is presented to extend the measurements of photon-photon
scattering  in ultra-peripheral Pb-Pb collisions at the LHC into
the mass region of the pseudoscalar resonances $\eta$ and $\eta'$.
The elementary photon-photon scattering cross section is presented.
The cross section for photon-photon scattering in Pb-Pb is derived
by convoluting the elementary photon-photon cross section with
the Pb-Pb photon luminosity. The main background to two-photon final states,
arising from double $\pi^{0}$ production with two of the four decay photons
escaping detection, is examined, and possible kinematical conditions are
discussed to optimize the signal-to-background ratio for such measurements
at mid-rapidity.

\end{abstract}
  
\section{Introduction}

Classical electrodynamics is epitomised by Maxwell's equations
\begin{equation}
  \partial_{\alpha}F^{\alpha\beta} = \frac{4\pi}{c}J^{\beta},\hspace{1.cm}
  \partial_{\alpha}{\cal F}^{\alpha\beta} = 0 \, .
\label{Eq:Maxwell}
\end{equation}

The Maxwell equations in vacuum are linear in the electromagnetic
fields {\bf E} and {\bf B}. Two electromagnetic waves will pass through
each other without scattering. The superposition principle conveniently
expresses the non-interaction of electromagnetic fields at the classical
level. The electromagnetic field energy carried by the electron, however,
poses a conceptual challenge at the classical level. Attempts to circumvent
the infinite Coulomb energy of a point charge resulted in Born-Infeld
electrodynamics which affirms the linearity of Maxwell's equations down
to some length scale intrinsic to the electron, and introduces non-linear
equations for smaller lengths scales~\cite{Born}.  The polarisation
of the vacuum in view of Dirac's positron theory led to the
Euler-Kockel-Heisenberg Lagrangian which modifies the classical Maxwell's
equation in vacuum by leading non-linear terms~\cite{Heisenberg}. The advent of
accelerating heavy-ions at the LHC has opened up the possibility of measuring
the photon-photon scattering cross section due to the large associated photon
luminosity of the heavy-ion beams. Evidence of such events have been reported by
the ATLAS and CMS collaborations at the LHC~\cite{ATLAS,CMS}. These measurements
are, however, restricted to photon-photon invariant masses W$_{\gamma\gamma} >$
5 and 6 GeV  for the CMS and ATLAS data, respectively. The purpose of the
analysis presented here is to study the feasibility of measuring photon-photon
scattering in the range 0.4$<$W$_{\gamma\gamma}<$5 GeV. As a first step,
we analyse the
corresponding cross section in this range, and examine the background
which is dominated by $\pi^{0}$ pair production with only two of the four
decay photons being within the detector acceptance.

\section{The elementary photon-photon scattering cross section}

Different mechanisms contribute to the elementary
$\gamma\gamma\rightarrow\gamma\gamma$ scattering. 

\begin{figure}[htb]
\centerline{%
\includegraphics[width=3.2cm]{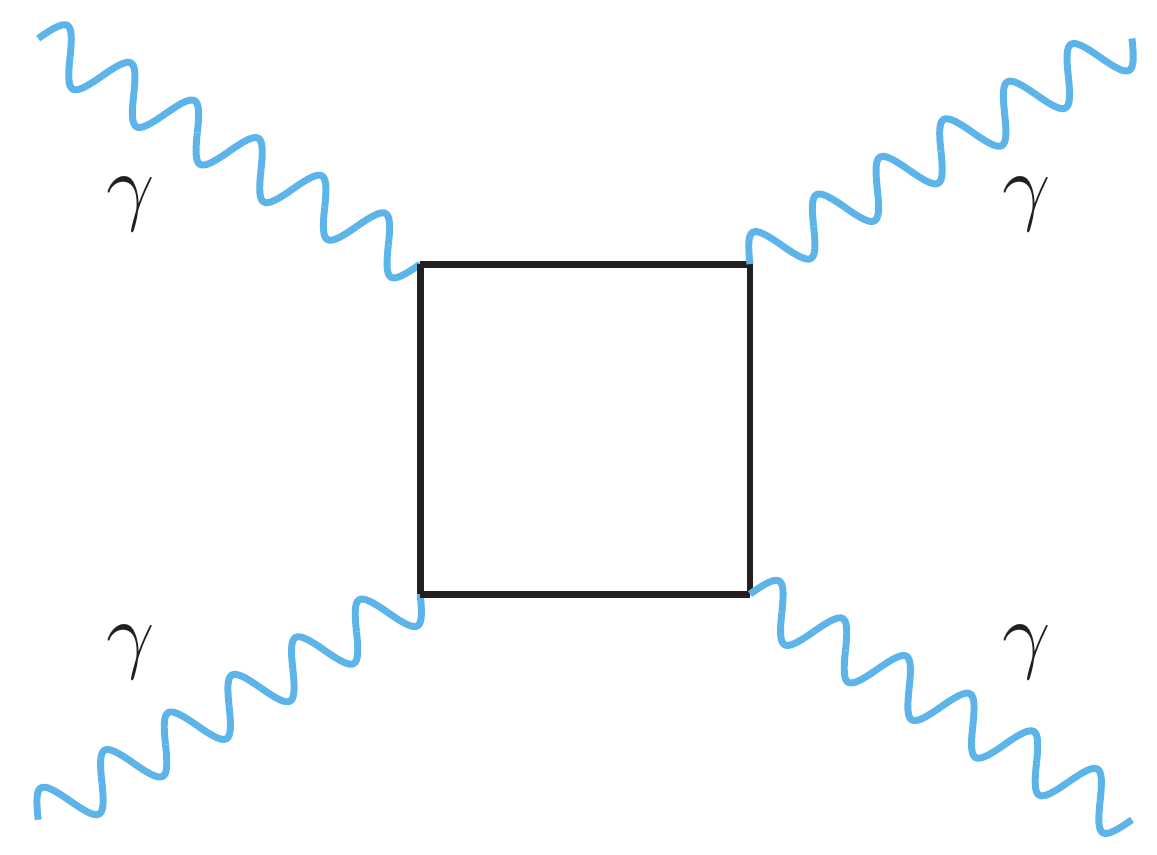}
\hspace{0.2cm}
\includegraphics[width=4.4cm]{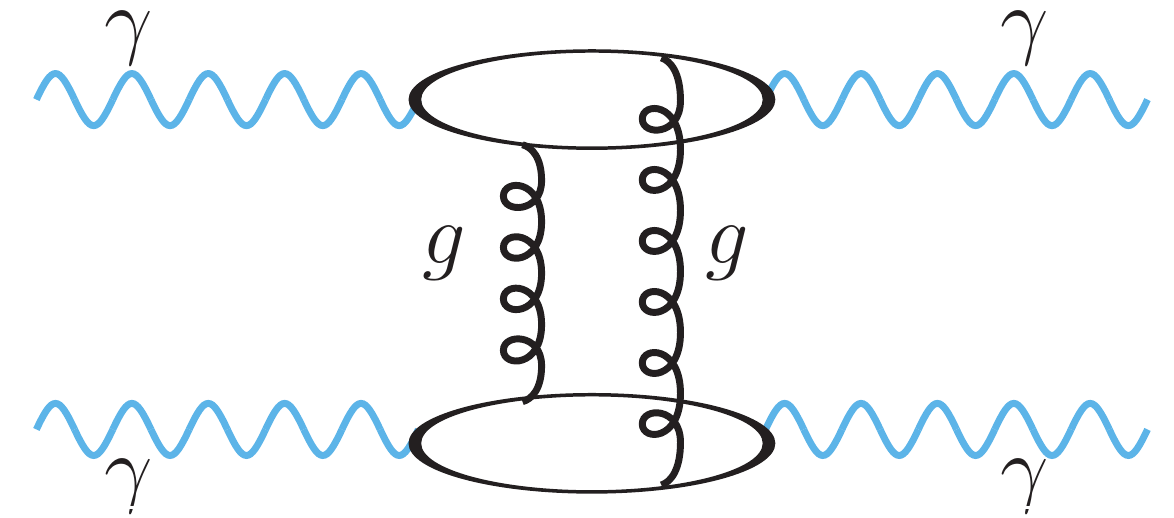}
\hspace{0.2cm}
\includegraphics[width=4.4cm]{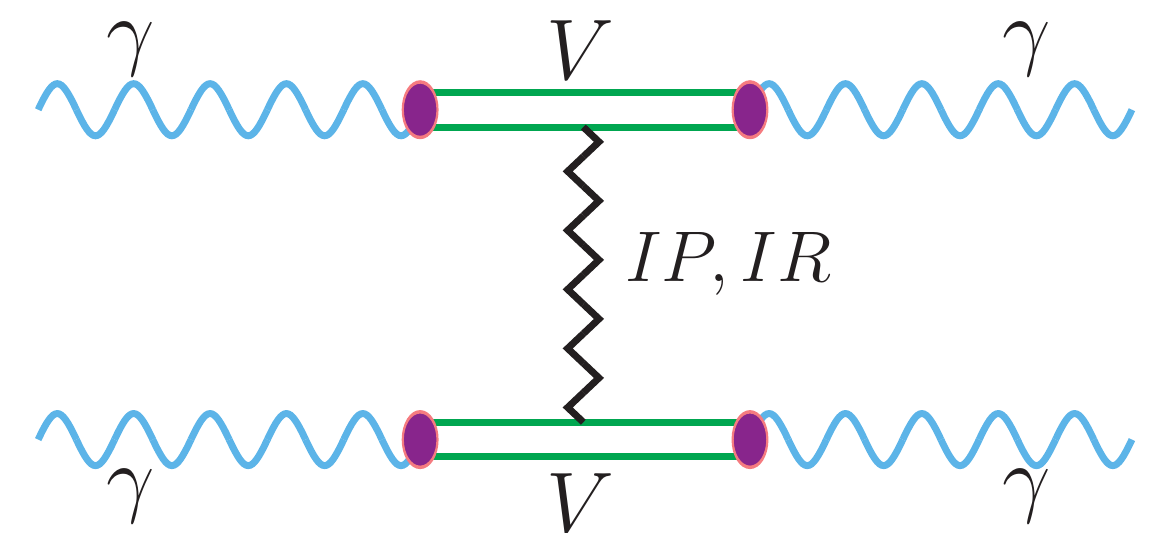}}
\caption{Mechanisms of $\gamma\gamma\rightarrow\gamma\gamma$ scattering
  (Fig. taken from Ref. \cite{Sz1}).}
\label{Fig:gg_elem}
\end{figure}

In Fig. \ref{Fig:gg_elem}, the different mechanisms of $\gamma\gamma$
scattering are presented. On the left, the loop diagram for fermions, leptons
and quarks, is shown. In the center, a QCD correction is displayed corresponding
to a three-loop mechanism. On the right, the analogous process as expressed
in the vector dominance approach is shown~\cite{Sz1}. 

\section{Photon-photon scattering in ultra-peripheral heavy-ion reactions}

The cross section for photon-photon scattering in ultra-peripheral heavy-ion
collisions can be calculated by folding the cross section of the elementary
mechanisms shown in Fig. \ref{Fig:gg_elem} with the equivalent photon
flux~\cite{Baur},
\begin{equation}
N(\omega,b) = \frac{Z^{2}\alpha}{\pi^{2}}
\frac{1}{\beta^{2}b^{2}}
\bigg| \int_0^{\infty} dv\; v^{2} J_{1}(v)\frac{F_{el}(-\frac{u^{2}+v^{2}}{b^{2}})}{u^{2}+v^{2}}\bigg|^{2} \, .
\label{Eq:EPA}
\end{equation}

The equivalent photon flux is defined in Eq. \ref{Eq:EPA}.
Here, $\omega$ denotes the energy of the photon, and b represents the
transverse distance from the center of the nucleus where the photon density
is evaluated. The integral in Eq. \ref{Eq:EPA} represents the form factor
of the charge distribution of the source.

\begin{figure}[htb]
\centerline{%
\includegraphics[width=6.5cm]{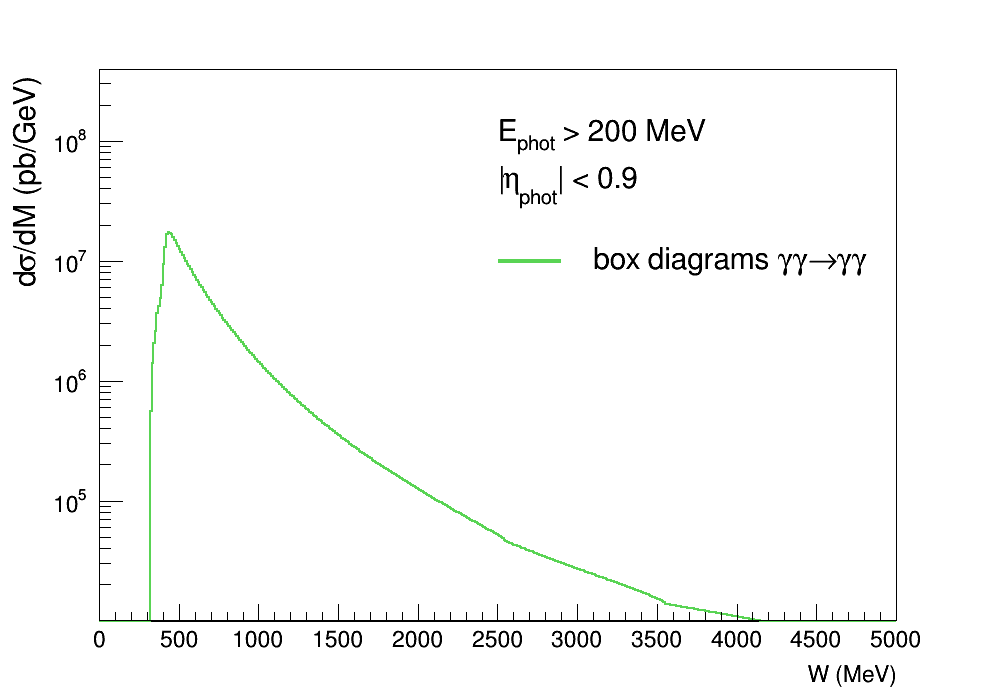}}
\caption{Differential cross section d$\sigma$/dM for photon scattering in PbPb-collisions.}
\label{Fig:dsigma1}
\end{figure}

The cross section for $\gamma\gamma\rightarrow\gamma\gamma$ in PbPb-collisions
is calculated by convoluting the elementary cross section with the
photon flux \mbox{of Eq. \ref{Eq:EPA}},

\begin{equation}
\sigma^{EPA}_{PbPb \rightarrow PbPb\gamma\gamma} = \int\!\!\int dn^{1}_{\gamma} 
\; dn^{2}_{\gamma} \; \sigma_{\gamma\gamma \rightarrow \gamma\gamma}(\omega_1,\omega_2) \, .
\label{Eq:dsigmadM_PbPb}
\end{equation}

The differential cross section for photon-photon scattering in PbPb-collisions
at $\sqrt{s}$=5.02 TeV resulting from the convolution of
Eq. \ref{Eq:dsigmadM_PbPb} is shown in Fig. \ref{Fig:dsigma1}.
This cross section is derived with the box diagrams of Fig. \ref{Fig:gg_elem},
and with conditions of the two final state photons being within the
pseudorapidity range $|\eta|<$0.9, and having an energy E$_{phot} >$200 MeV.

\section{Resonance signal from $\eta,\eta'$ decays}

The cross section for photoproduction of $\eta,\eta'$ is taken
according to \cite{Budnev}
\begin{equation}
  \sigma_{\gamma\gamma\rightarrow R} = 8\pi(2J+1)\frac{\Gamma_{\gamma\gamma}\Gamma_{tot}}{(W^{2}-M^{2}_{R})^{2}+M^{2}_{R}\Gamma^{2}_{tot}} \, .
  \label{Eq:sigma_res}
\end{equation}

The cross section for photoproduction of $\eta,\eta'$ in
PbPb-collisions is calculated according to the convolution defined
in Eq. \ref{Eq:dsigmadM_PbPb}.
  
\begin{figure}[htb]
\centerline{%
\includegraphics[width=6.5cm]{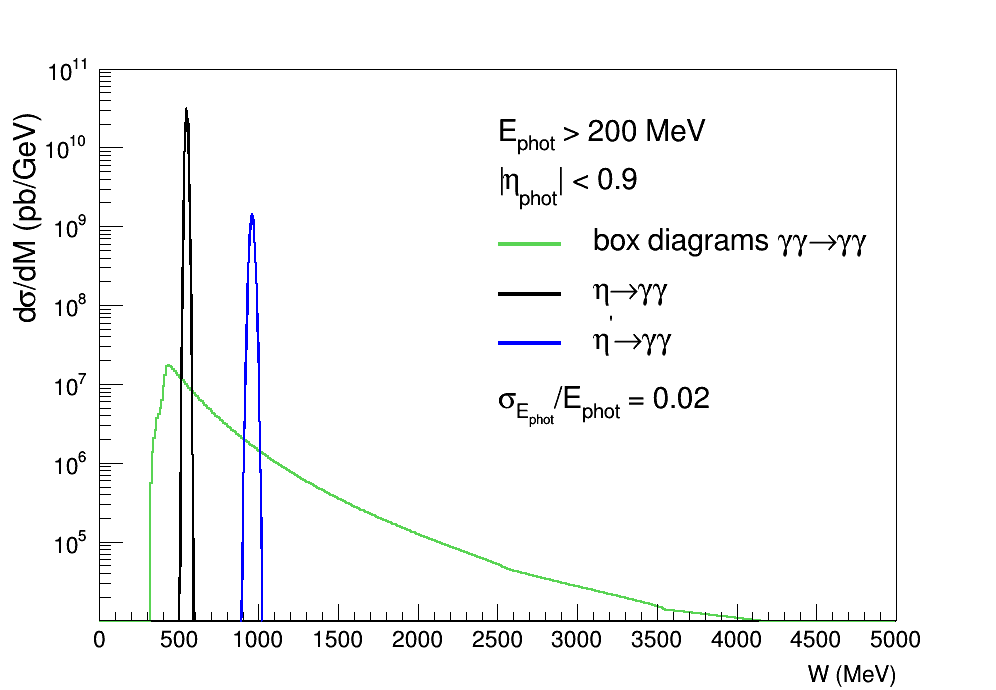}}
\caption{Cross section for photoproduction of $\eta,\eta'.$}
\label{Fig:dsigmadM_eta}
\end{figure}

The $\eta,\eta'$ cross section in PbPb-collisions at $\sqrt{s}$= 5.02 TeV
multiplied by the branching ratio $\eta,\eta'\rightarrow\gamma\gamma$ is shown
in Fig. \ref{Fig:dsigmadM_eta}. The values shown are derived with conditions
of the two decay photons being within the pseudorapidity range $|\eta|<$0.9,
and having an energy E$_{phot} >$200 MeV. The finite width of these two
resonances results from the detector resolution of the photon
measurements which is taken here to be $\sigma_{E_{phot}}/E$  = 0.02.

\section{Photoproduction of $\pi^{0}\pi^{0}$ pairs}

The main background to the signal of two photons in the final state
results from $\pi^{0}\pi^{0}$ production with two of the four decay photons 
escaping detection. The two measured photons from this $\pi^{0}\pi^{0}$ decay
cannot be distinguished from the signal of photon-photon scattering.

\begin{figure}[htb]
\centerline{%
\includegraphics[width=4.cm]{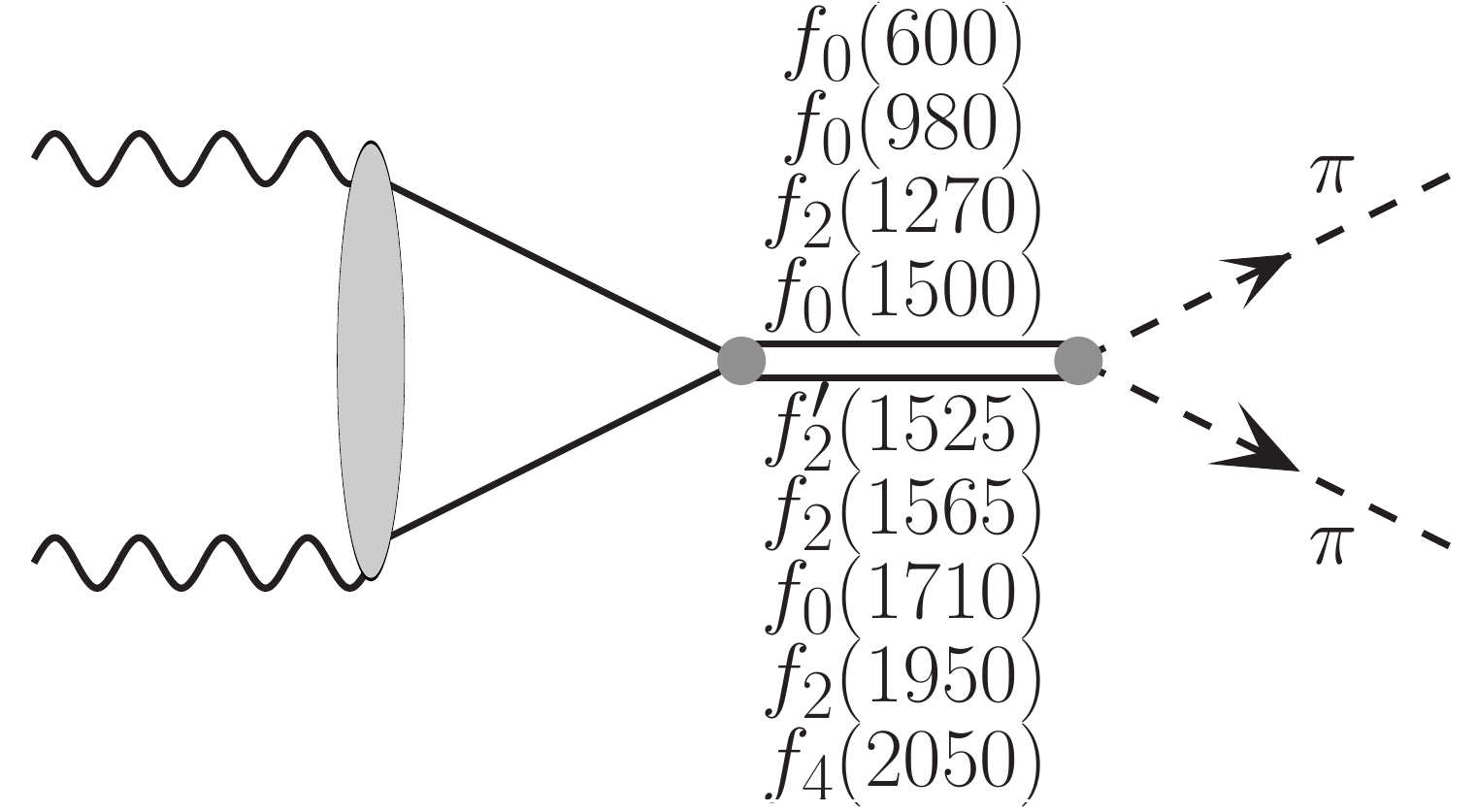}
\hspace{0.3cm}
\includegraphics[width=4.cm]{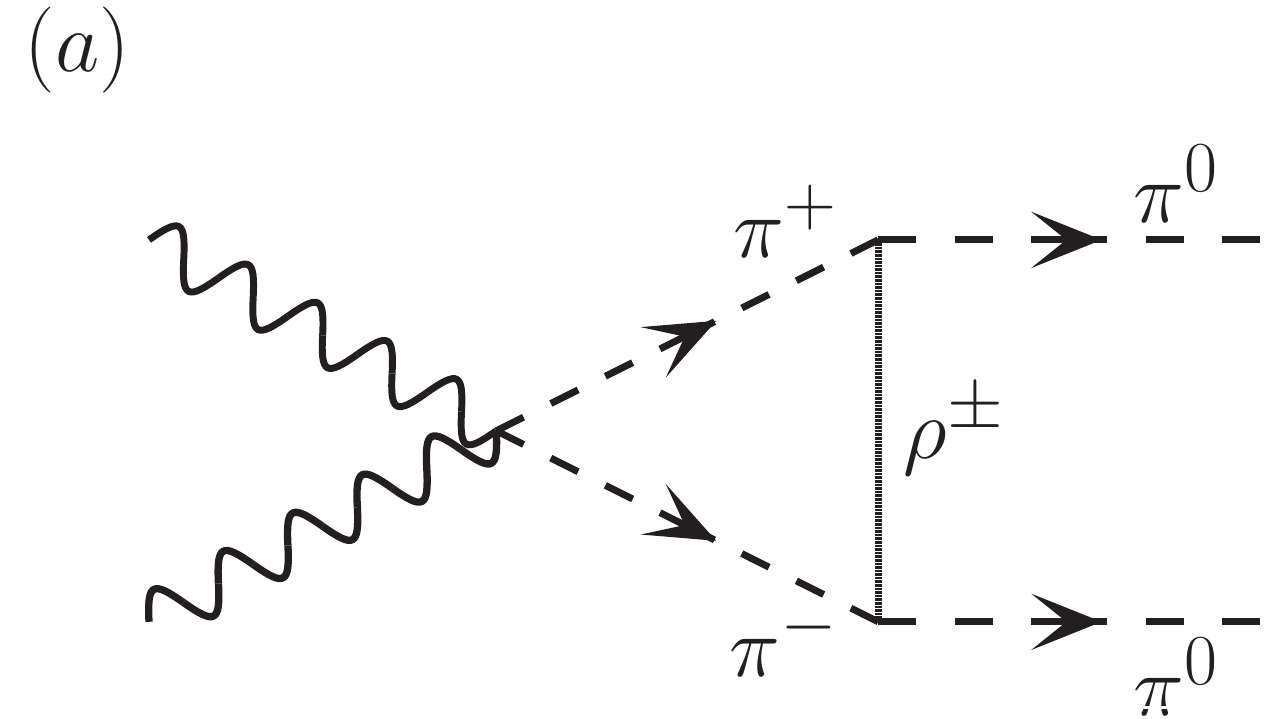}
\hspace{0.3cm}
\includegraphics[width=4.cm]{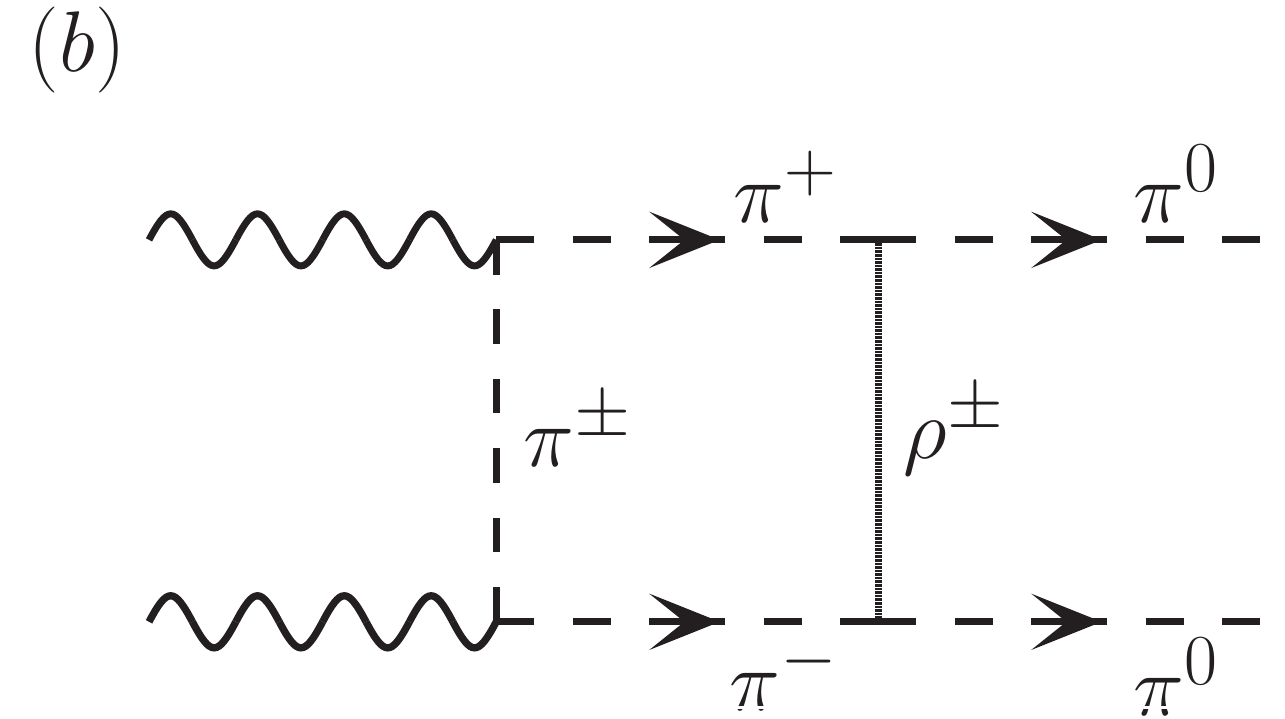}}
\caption{Mechanisms of $\pi^{0}\pi^{0}$ photoproduction
  (Fig. taken from \cite{Sz2}).}
\label{Fig:pion_pairs}
\end{figure}

The contribution of resonance decays to the $\pi^{0}\pi^{0}$ final state is
shown in Fig. \ref{Fig:pion_pairs}  on the left. Here, the resonances
$\sigma$(600), $f_{0}$(980), $f_{0}$(1500), $f_{0}$(1710), $f_{2}$(1270),
$f^{`}_{2}$(1525), $f_{2}$(1565), $f_{2}$(1950) and $f_{4}$(2050) are
considered~\cite{Sz2}. The cross section $\gamma\gamma\rightarrow\pi^{+}\pi^{-}$
is much larger than for  $\gamma\gamma\rightarrow\pi^{0}\pi^{0}$, hence even
a small coupling between the charged and neutral pion channel might have an
influence on the $\gamma\gamma\rightarrow\pi^{0}\pi^{0}$ cross section.
Examples of processes leading to such channel couplings are shown in
Fig. \ref{Fig:pion_pairs} in the middle and on the right.

\section{Background from $\pi^{0}\pi^{0}$ decays}

\begin{figure}[htb]
\centerline{%
\includegraphics[width=6.5cm]{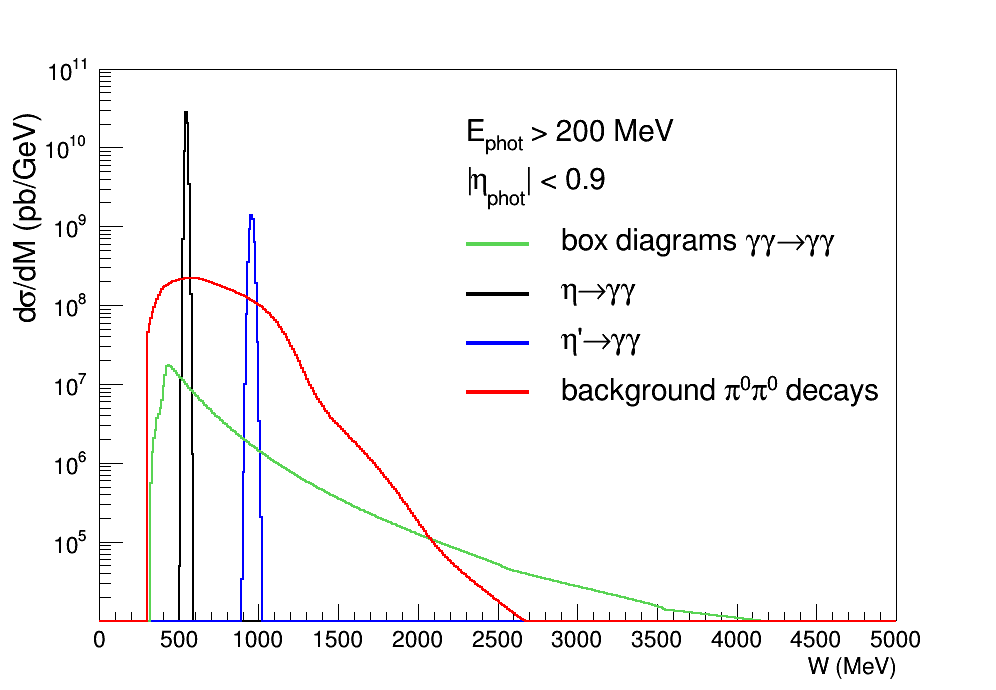}}
\caption{Signal and background from $\pi^{0}\pi^{0}$ decays.}
\label{Fig:bck_pi0}
\end{figure}

The $\pi^{0}\pi^{0}$ cross section in PbPb-collisions is calculated
\mbox{by convoluting} the elementary $\pi^{0}\pi^{0}$ cross section
with the photon flux according to Eq. \ref{Eq:EPA}.
The background from $\pi^{0}\pi^{0}$ decays is shown in Fig. \ref{Fig:bck_pi0}
by the solid red line. This background cross section is derived by the
conditions that exactly one decay photon from each of the two
$\pi^{0}$'s is within the pseudorapidity range $|\eta| <$0.9, and that
these photons have an energy E$_{phot} >$200 MeV.
\vspace{-.4cm}

\subsection{Background suppression by asymmetry cuts}

The two photons of the signal are of equal transverse momentum and are
back-to-back in azimuth. These correlations are smeared out due to finite
resolution in the measurement of photon energy and azimuthal angle. The
two photons from $\pi^{0}\pi^{0}$ decay do not show these correlations.
\mbox{A scalar ($A_{S}$)} and vector ($A_{V}$) asymmetry can be defined for
background suppression,
\begin{equation}
  A_S=\left|\frac{|\vec{p}_T(1)|-|\vec{p}_T(2)|}
  {|\vec{p}_T(1)|+|\vec{p}_T(2)|}\right|,\hspace{1.cm}
A_V=\frac{|\vec{p}_T(1)-\vec{p}_T(2)|}
{|\vec{p}_T(1)+\vec{p}_T(2)|} \, .
\end{equation}
\vspace{-.8cm}
\begin{figure}[htb]
\centerline{%
\includegraphics[width=6.5cm]{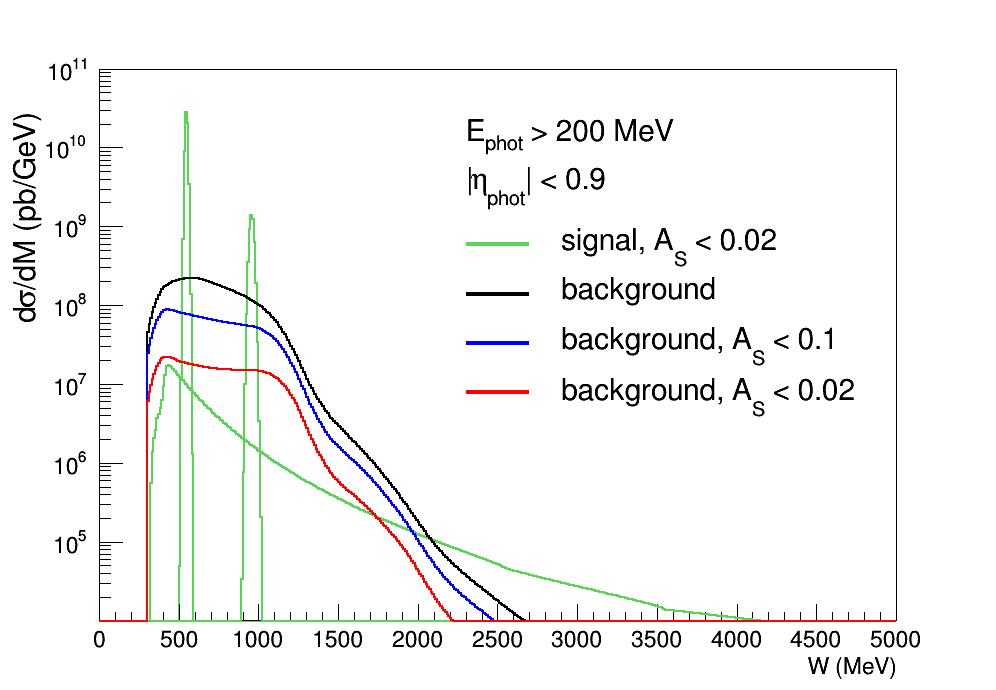}}
\caption{Background suppression by asymmetry cut A$_{S}$.}
\label{Fig:bck_pi0_asym}
\end{figure}

The background reduction  for condition A$_{S}\!<$0.1 and
A$_{S}\!<$0.02 is shown in Fig. \ref{Fig:bck_pi0_asym} by the blue and red
line, respectively. The condition A$_{S}<$0.02 reduces the background by
about a factor $\sim$10 while keeping 98\% \mbox{of the signal.} 

\subsection{Background correction by sideband subtraction}

The background remaining after the cut A$_{S}\!<\!0.02$
can be subtracted by  sideband correction. The signal band is given
by \mbox{$0.0\!<\!A_{s}\!<\!0.02$,} with sideband 1 and 2 defined by
$0.02\!<\!A_{s}\!<\!0.04$ and  $0.04\!<\!A_{s}\!<\!0.06$, respectively.
An estimator for the background in the signal region can be defined
by linear extrapolation of the sidebands 1 and 2 into the signal region.

\vspace{-.4cm}
\begin{figure}[htb]
\centerline{%
\includegraphics[width=6.5cm]{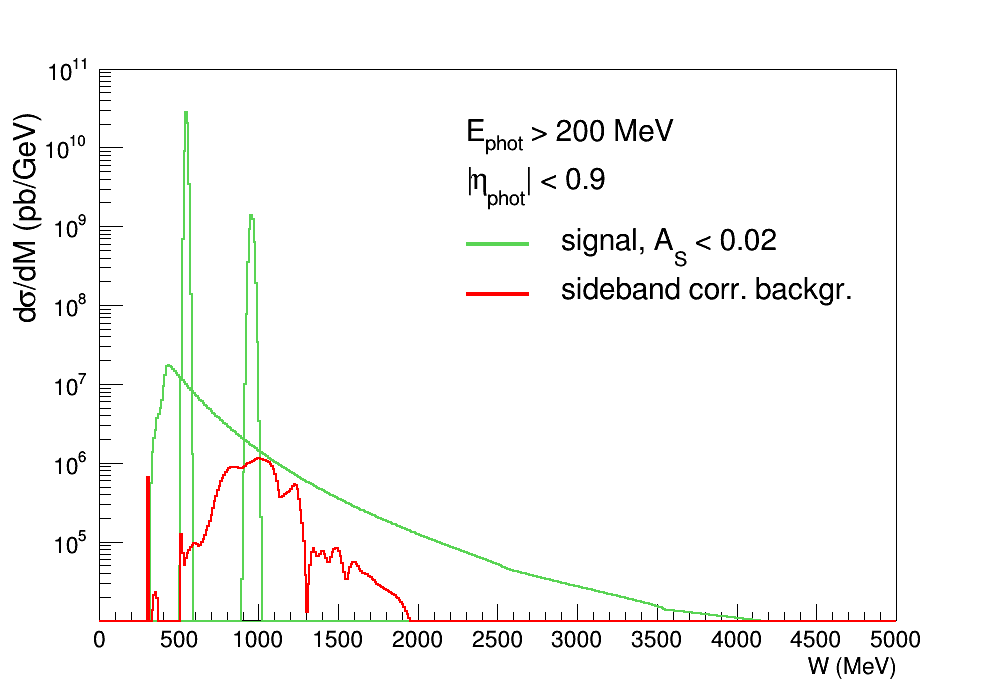}}
\caption{Sideband corrected background.}
\label{Fig:bck_pi0_asym_corr}
\end{figure}
\vspace{-0.2cm}
The background remaining after sideband correction is shown
\mbox{in Fig. \ref{Fig:bck_pi0_asym_corr}} by the red line. The sideband
correction could be further improved by defining more than two sidebands,
and by a non-linear extrapolation of the background into the signal region.

\vspace{-0.2cm}
\section{ACKNOWLEDGMENTS}
This work is supported by the German Federal Ministry of Education and 
Research under promotional reference 05P15VHCA1.


\vspace{-0.6cm}
\begin{thebibliography}{99}
\vspace{-0.3cm}
\bibitem{Born}M.Born, Nature 132:3329 (1933) 282.
\bibitem{Heisenberg}W.Heisenberg, H.Euler, Zeitschrift f\"{u}r Physik 98:
  11-12 (1936) 714.
\bibitem{ATLAS}ATLAS Collaboration, Nature Phys. 13 (2017) no.9, 852.
\bibitem{CMS}CMS Collaboration, arXiv:1810.04602.
\bibitem{Sz1}M.K\l{}usek-Gawenda et al., Phys.Rev. C93 (2016), no.4, 044907. 
\bibitem{Baur}G.Baur et al., Phys.Rept. 364 (2002) 359.
\bibitem{Budnev}V.M.Budnev, $\!$I.F.Ginzburg, $\!$G.V.Meledin, $\!$V.G.Serbo,
  $\!$Phys.Rept.15, $\!$(1975) $\!$181.
\bibitem{Sz2}M.K\l{}usek-Gawenda, A.Szczurek, Phys.Rev. C87 (2013), no.5, 054908. 
\end{thebibliography}
\end{document}